\documentstyle[prl,aps,twocolumn,float,epsfig]{revtex}

\def\beq{\begin{equation}}
\def\eeq{\end{equation}}
\def\beqa{\begin{eqnarray}}
\def\eeqa{\end{eqnarray}}
\def\bfig{\begin{figure}}
\def\efig{\end{figure}}

\begin{document}
\draft
\fnsymbol{footnote}

\wideabs{

\title{Nonlinear Evolution of the $r$-Modes in Neutron Stars
%\footnotemark
}

\author{Lee Lindblom${}^1$, Joel E. Tohline${}^2$, and Michele 
Vallisneri${}^{1,3}$}
\address{${}^1$Theoretical Astrophysics 130-33,
         California Institute of Technology,
         Pasadena, CA 91125}

\address{${}^2$Department of Physics and Astronomy, 
Louisiana State University, Baton Rouge, LA 70803}

\address{${}^3$INFN, Sezione di Milano, Gruppo Collegato di Parma/Universit\`a
di Parma, 43100 Parma, Italy}

\date{\today}
\maketitle

\begin{abstract} The evolution of a neutron-star $r$-mode 
driven unstable by gravitational radiation (GR) is studied here using
numerical solutions of the full nonlinear fluid equations.  The
dimensionless amplitude of the mode grows to order unity before strong
shocks develop which quickly damp the mode.  In this simulation the
star loses about 40\% of its initial angular momentum and 50\% of its
rotational kinetic energy before the mode is damped.  The nonlinear
evolution causes the fluid to develop strong differential rotation
which is concentrated near the surface and poles of the star.
\end{abstract}
}
%%%%%%%%%%%%%%%%%%%%%%%%%%%%%%%%%%%%%%%%%%%%%%%%%%%%%%%%%%%%%%%%%%%%%%%%%%%%%%%
The $r$-modes of all rotating stars are driven towards instability by
gravitational radiation (GR) reaction~\cite{andersson,fm}, and the
strength of this destabilizing force is sufficient to dominate over
internal dissipation in hot, rapidly rotating neutron
stars~\cite{lom}.  The growth timescale of the instability is about 40
s for neutron stars with millisecond rotation periods.  Thus it is
generally expected that GR will cause the dimensionless amplitude of
the most unstable ($m=2$) $r$-mode to grow to order unity within about
ten minutes of the birth of such a star.  The emission of GR by this
process removes angular momentum and rotational kinetic energy from
the star.  The strength of the GR emitted and the timescale for
spinning down the young neutron star depend critically on the
amplitude to which the $r$-mode grows.  Initial estimates assumed that
the amplitude would grow to order unity before some unknown process
would saturate the mode.  With this saturation amplitude, a neutron
star spins down to about one tenth its maximum angular velocity in
about one year, and the GR from this event might be detectable by LIGO
II~\cite{owen}.  Stergioulas and Font~\cite{stergioulas} find no
saturation of the $r$-modes even at large amplitudes in their
nonlinear numerical study.  Thus at present no one really knows what
process will saturate the $r$-modes, or how large their amplitudes
will grow.  Here we investigate the growth of the $r$-modes by solving
numerically the nonlinear hydrodynamic equations driven by GR
reaction.

Neutron stars are compact objects with reasonably strong gravitational
fields, $GM/R\approx 0.2c^2$.  While these objects may contain fluid
moving at fairly large velocities, $v^2\lesssim GM/R$, Newtonian
theory is expected to describe them up to errors of order
$GM/Rc^2$.  For simplicity then we study here the
nonlinear evolution of the $r$-modes using the Newtonian
equations:

\beq
\partial_t\rho+\vec{\nabla}\cdot\bigl(\rho\vec{v}\bigr)=0,\label{eq1}
\eeq
 
\beq
\rho\bigl(\partial_t\vec{v}+\vec{v}\cdot\vec{\nabla}\vec{v}\bigr) =
-\vec{\nabla}p-\rho\vec\nabla\Phi 
+ \rho\vec{F}_{GR},
\label{eq2}
\eeq

\noindent where $\vec{v}$ is the fluid velocity,
$\rho$ and $p$ are the density and pressure, $\Phi$ is the Newtonian
gravitational potential, and $\vec{F}_{GR}$ is the GR reaction force.
The gravitational potential is determined by Poisson's equation,
                                                     
\beq
\nabla^2\Phi = 4\pi G\rho.\label{eq3}
\eeq

\noindent The GR reaction force $\vec F_{GR}$ due to a time-varying 
current quadrupole (the dominant multipole for the $r$-modes) can be
written (see Blanchet~\cite{blanchet} and Rezzolla, {\it
et al.}~\cite{rezzolla}):

\beqa
&&F_{GR}^x-iF_{GR}^y = -
\kappa i (x+iy)\Bigl[3v^z J^{(5)}_{22} + z J^{(6)}_{22}\Bigr],\label{eq4}\\
&&F_{GR}^z =-\kappa \,{\rm Im}
\Bigl\{(x+iy)^2\Bigl[3{v^x+iv^y\over x+iy}J^{(5)}_{22}+J^{(6)}_{22}\Bigr]
\Bigr\}, 
\label{eq5}
\eeqa

\noindent where $J^{(n)}_{22}$ represents the $n$'th time 
derivative of $J_{22}$,

\beq
J_{22} = \int \rho r^{2} \vec{v}\cdot\vec{Y}^{B*}_{22}d^{\,3}\!x,
\label{eq6}
\eeq

\noindent and $\vec Y^B_{22}=\hat r \times r\vec
\nabla Y_{22}/\sqrt{6}$ is the magnetic-type vector spherical harmonic.  
In slowly rotating stars the $m=2$ $r$-mode projects onto $J_{22}$ and
no other $J_{lm}$.  The parameter $\kappa$ that appears in
Eqs.~(\ref{eq4}) and (\ref{eq5}) sets the strength of the GR reaction
force, and has the value $\kappa=32\sqrt{\pi}G/(45\sqrt{5}c^7)$ in
general relativity theory.  For practical reasons (see further
discussion below), we take $\kappa$ to be about 4500 times this
value.

We solve Eqs.~(\ref{eq1})--(\ref{eq3})
numerically in a rotating reference frame using the computational
algorithm developed at LSU to study a variety of astrophysical
hydrodynamic problems~\cite{tohline}.  Briefly, the code performs an
explicit time integration of the equations using a finite-difference
technique that is accurate to second order both in space and time, and
uses techniques very similar to those of the familiar ZEUS
code~\cite{stone_norman}.

We find that as written, Eqs.~(\ref{eq4}) and (\ref{eq5}) for
$\vec{F}_{GR}$ are nearly useless in a numerical evolution.  The
problem is the large number of time derivatives of $J_{22}$ that
appear there.  In the case of a pure mode with frequency $\omega$ this
problem is easily solved: $J^{(n)}_{22}=(i\omega)^n J_{22}$.  Even
when the amplitude of the $r$-mode becomes large, we expect the fluid
motion to be dominated by periodic motions at the fundamental
frequency of the $r$-mode.  Thus, we expect the normal-mode
expressions for the time derivatives of the multipole moments to be
reasonably accurate even in the nonlinear regime.  It is easy to
evaluate $J_{22}$ from Eq.~(\ref{eq6}), and $J^{(1)}_{22}$ can also be
expressed as an integral over the fluid variables using
Eqs.~(\ref{eq1}) and (\ref{eq2}):

\beq
{J}^{(1)}_{22} = \int\rho\Bigl[
\vec{v}\cdot\vec{\nabla}\bigl(r^2\vec{Y}^{B*}_{22}\bigr)\cdot\vec{v} -
r^2\vec{\nabla}\Phi\cdot \vec{Y}^{B*}_{22}\Bigr]
d^{\,3}\!x.
\label{eq7}
\eeq

\noindent Thus we evaluate the time derivatives needed in 
Eqs.~(\ref{eq4}) and (\ref{eq5}) using $J^{(5)}_{22}=\omega^4
J^{(1)}_{22}$ and $J^{(6)}_{22}=-\omega^6 J_{22}$.  We have verified
that using these approximations our numerical code accurately
reproduces the analytical description of the $r$-mode evolution and
growth due to GR reaction in slowly rotating models.

In order to monitor the nonlinear evolution of an $r$-mode, it will
be helpful to introduce nonlinear generalizations of the amplitude
and frequency of the mode.  Since the $r$-mode projects primarily
onto the multipole moment $J_{22}$, we define the nonlinear
amplitude to be

\beq
\alpha = \frac{8 \pi R_0|J_{22}|}{\Omega_0\int \rho r^4 d^{\,3}\!x}
,\label{eq8}
\eeq

\noindent where $\Omega_0$ is the initial angular velocity and $R_0$ is the radius of the corresponding nonrotating stellar model.
This $\alpha$ is normalized to reduce to the standard definition used
in perturbations of slowly rotating stars \cite{lom}.  We must also
define a generalization of the frequency of the $r$-mode.  For a
small-amplitude mode the time derivative of $J_{22}$ is proportional
to the frequency: ${J}^{(1)}_{22} = i\omega J_{22}$.  Thus we are led
to define the following nonlinear generalization of the $r$-mode
frequency

\beq
\omega = - \frac{|{J}^{(1)}_{22}|}{|J_{22}|}.\label{eq9}
\eeq

\noindent These expressions for $\alpha$ and $\omega$ are very 
stable numerically since they are expressed as integrals.~\cite{note1}

We have studied the growth of an $r$-mode as described above by
solving numerically the nonlinear hydrodynamic equations for a
rotating stellar model represented on a $64\times 128\times 128$
cylindrical grid.  We prepare initial data for this evolution by
constructing first a rigidly rotating equilibrium stellar model.  For
simplicity we use the polytropic equation of state $p=K\rho^2$, which
approximates the features of more realistic neutron-star models
reasonably well.  The model most extensively studied here is very
rapidly rotating, with angular velocity $\Omega_0= 0.635\sqrt{\pi G
\bar\rho_0}$ where $\bar\rho_0$ is the average density of the
nonrotating star of the same mass.  The ratio of the initial
rotational kinetic energy to gravitational potential energy for this
model is $T/|W|=0.101$.  At the initial time we take the fluid density
to be that of this equilibrium stellar model, while the fluid velocity
is taken to include a small amplitude $r$-mode perturbation:

\beq
\vec{v} = \Omega_0\vec{\varphi} + \alpha_0 R_0\Omega_0
\left(\frac{r}{R_0}\right)^2 {\rm Re}(\vec{Y}^{B}_{22}).
\label{eq10}
\eeq

\noindent  In our simulation we take the initial $r$-mode amplitude
to be $\alpha_0=0.1$.  

According to perturbation theory~\cite{lom}, the ratio of the GR
growth time to the $r$-mode pulsation period is expected to be
$\tau_{GR}/P_r= 5.64\times 10^4$ for the stellar model studied here.
At this rate, growth of the $r$-mode is far too slow to be studied
effectively via an explicit hydrodynamic simulation.  To overcome
this, we artificially increase $\kappa$ in Eqs.~(\ref{eq4}) and
(\ref{eq5}) to a value such that $\tau_{GR}/P_r\approx12.6$, 
i.~e., the strength of our GR reaction is about 4500 times stronger
than it should be.  This made the growth rate of the $r$-mode large
enough so that it could be followed up to nonlinear amplitude with a
reasonable amount of computing time, yet kept it slow relative to the
dynamical and sound-crossing times.

\bfig
\centerline{\psfig{file=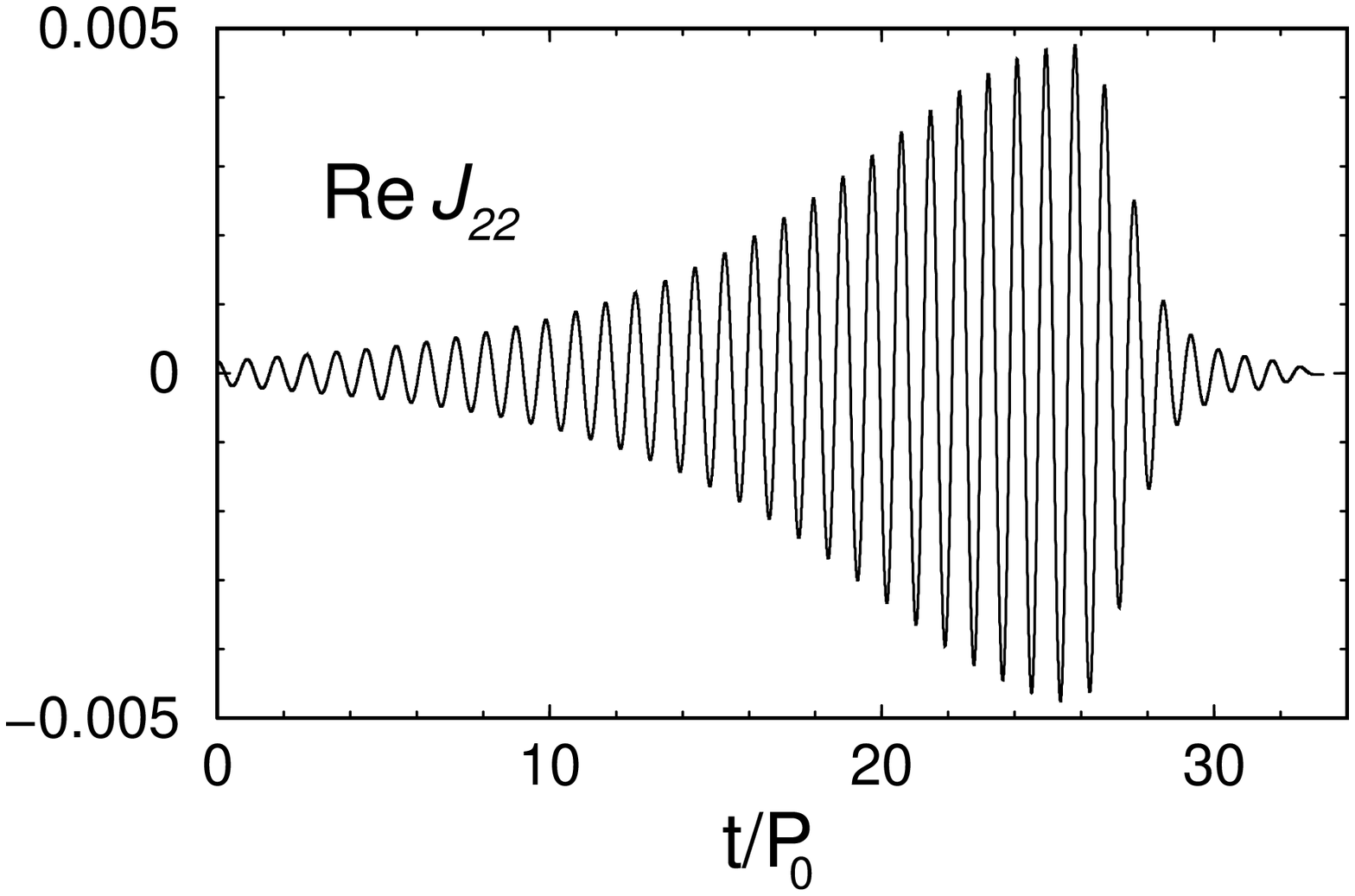,height=1.6in}}
\vskip 0.1cm
\caption{Evolution of the current quadrupole moment $J_{22}$.
\label{fig1}}
\efig
\bfig
\centerline{\psfig{file=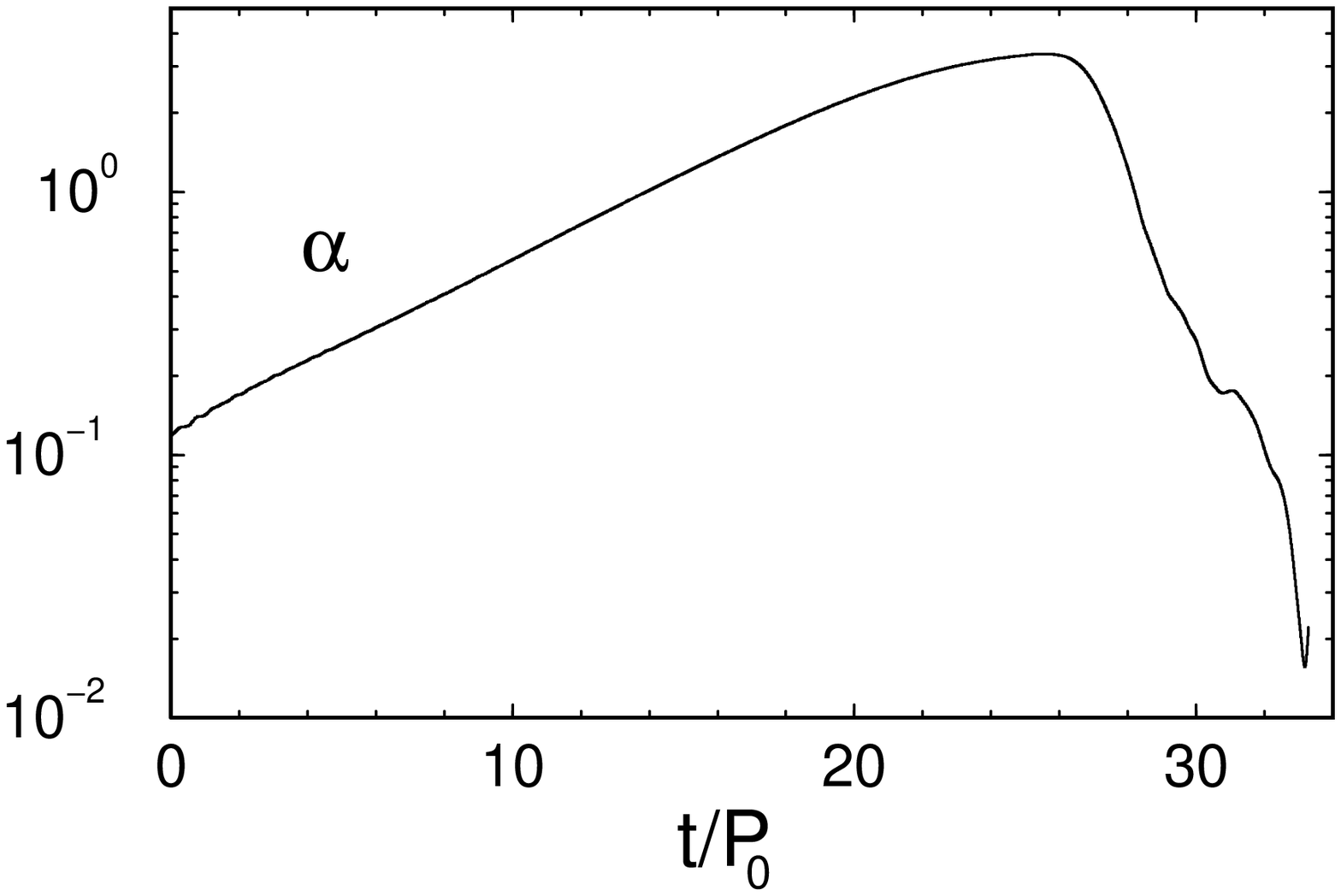,height=1.6in}}
\vskip 0.1cm
\caption{Nonlinear evolution of the $r$-mode amplitude $\alpha$.
\label{fig2}}
\efig

Figure~\ref{fig1} shows the ${\rm Re}(J_{22})$ that results from the
evolution of this system.  Time in these figures is given in units of
the initial rotation period of the star: $P_0=2\pi/\Omega_0$.
Figure~\ref{fig1} illustrates that the evolution of $J_{22}$ is
dominated by the sinusoidal $r$-mode oscillations.  Figure~\ref{fig2}
illustrates the evolution of the $r$-mode amplitude in this
simulation.  We see that the growth is exponential (as predicted by
perturbation theory) until $\alpha\approx 2$.  Then some nonlinear
process limits the growth; $\alpha$ peaks at $\alpha=3.35$ and then
decreases rapidly.  The evolution of the $r$-mode frequency $\omega$
defined by Eq.~(\ref{eq9}) is illustrated in Fig.~\ref{fig3}.  The
evolution of $\omega$ is quite smooth when the amplitude of the
$r$-mode is large: $\alpha\gtrsim0.5$.  At early ($t\lesssim 10P_0$)
and at late ($t\gtrsim28P_0$) times when the $r$-mode amplitude is
small, we see that other modes also make noticeable contributions to
$J_{22}$, and hence to $\omega$.

\bfig
\centerline{\psfig{file=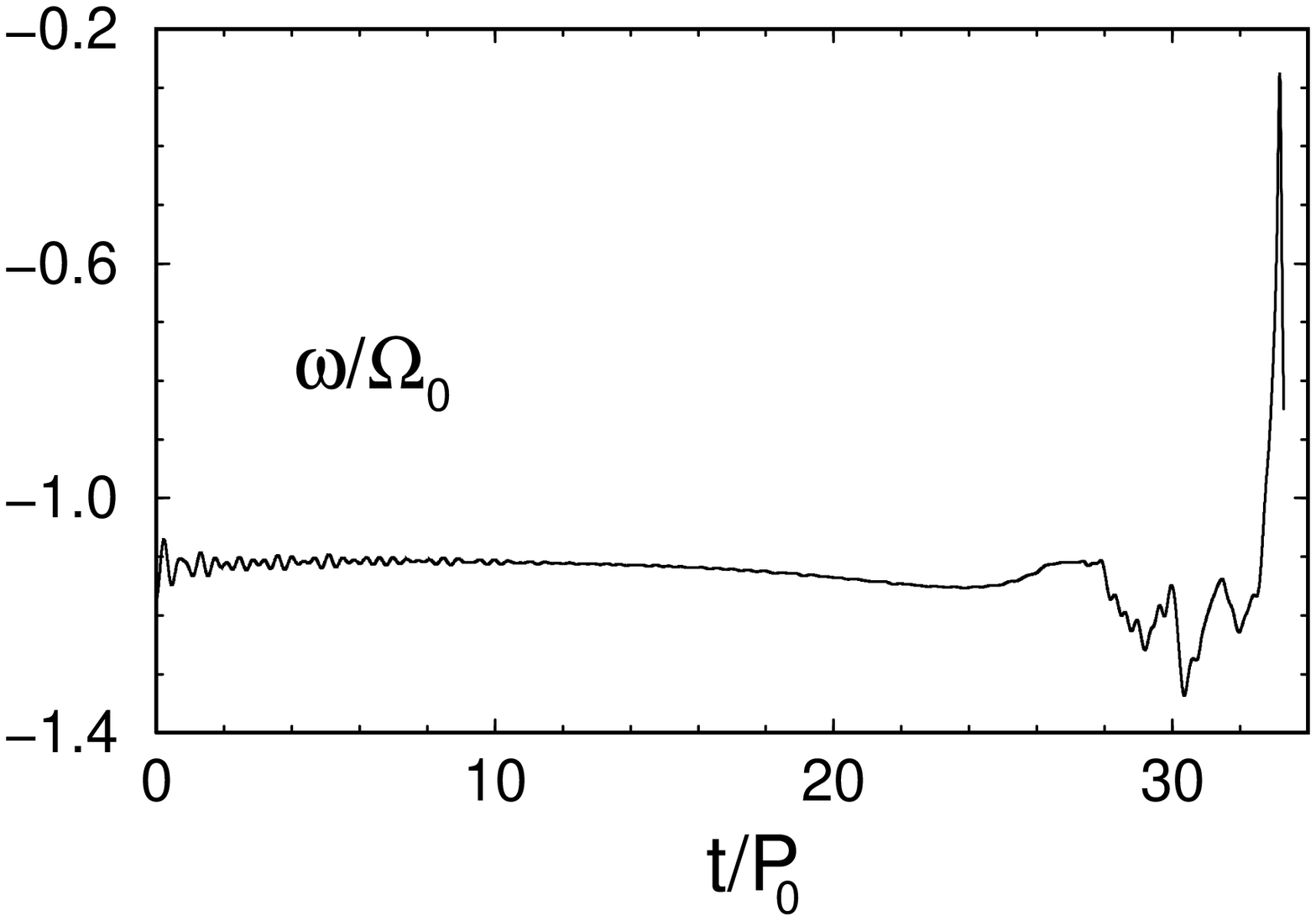,height=1.6in}}
\vskip 0.1cm
\caption{Evolution of the $r$-mode frequency.
\label{fig3}}
\efig

Several authors~\cite{diffrot} have suggested that the nonlinear
evolution of an unstable $r$-mode could generate differential
rotation.  This differential rotation would amplify existing magnetic
fields in the star, and these in turn might significantly affect the
evolution of the $r$-mode.  We have explored this possibility by
monitoring the average differential rotation $\Delta\Omega$.  We find
that $\Delta\Omega$ grows to $\Delta\Omega\approx0.41\bar\Omega$ at
time $t\approx 28P_0$ and then decreases.  Here $\bar\Omega$ is the
ratio of the angular momentum to the moment of inertia of the star.
But the average value of $\Delta\Omega/\bar\Omega$ may be misleading.
Figure~\ref{fig4} illustrates the spatial dependence of the
azimuthally averaged angular velocity $\Omega(\varpi,z)=\int\Omega
d\varphi/2\pi$ at the time $t=25.6P_0$.  We see that the differential
rotation is confined mostly to a thin shell of material near the
surface of the star, and is particularly concentrated near each polar
cap.  The bulk of the material in the star remains fairly rigidly
rotating.  Thus it appears that the magnetic fields generated by this
differential rotation may have strong effects locally (e.g., leading to
the ejection of material at the surface or along the rotation axis),
but may not affect the global behavior of the $r$-mode as much as if
the differential rotation were distributed more uniformly.

\bfig
\centerline{\hskip 0.5cm\psfig{file=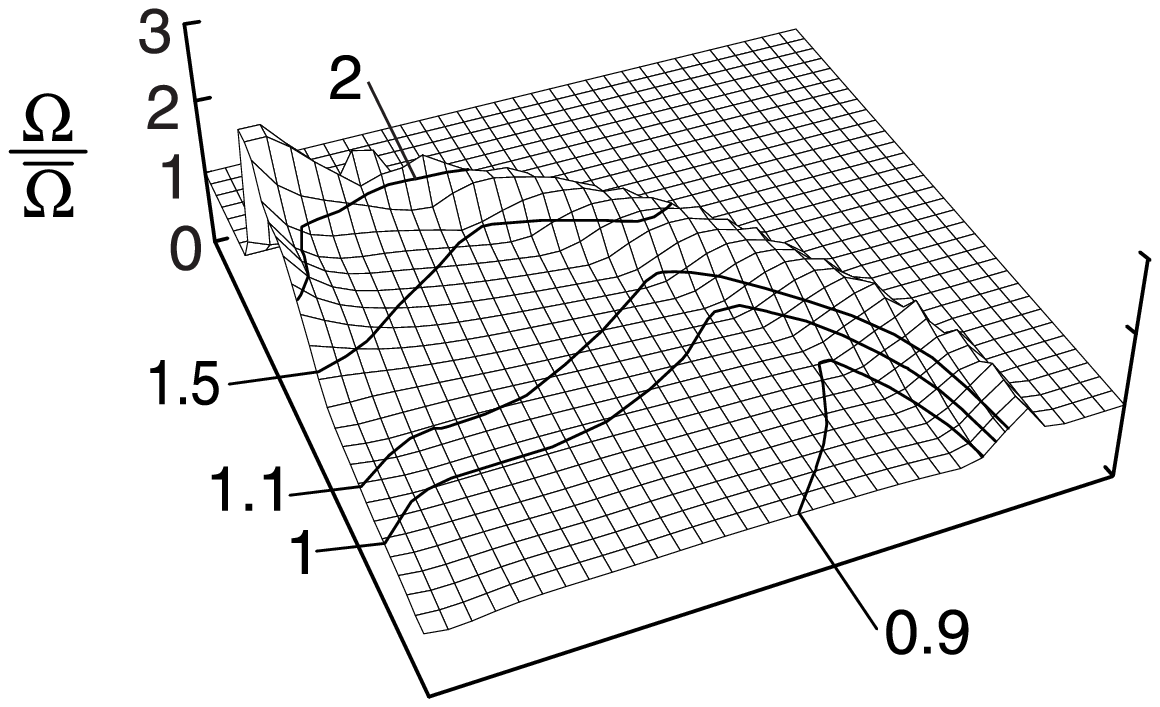,height=1.5in}}
\vskip 0.2cm
\caption{Spatial dependence of the azimuthally averaged angular velocity
$\Omega(\varpi,z)/\bar\Omega$.  Rotation axis is the left edge of the
figure; equatorial plane is the bottom edge.
\label{fig4}}
\efig

Energy and angular momentum are removed from the star by GR emission
from the $r$-mode according to the expression~\cite{rezzolla,thorne}:

\beq
{dE\over dt} = {|\omega|\over 2}{dJ\over dt}
= - {128\pi \over 225}{G \over c^7}\kappa  \omega^6 |J_{22}|^2.
\label{eq11}
\eeq

\noindent Figure~\ref{fig5} illustrates the evolution of the total angular
momentum $J$, the total mass $M$, and the kinetic energy of the fluid
$T$.  We see that $M$ is essentially unchanged, while $J$ decreases by
about 40\%.  The numerical evolution of $J$ agrees with
Eq.~(\ref{eq11}) to within a few percent~\cite{note3}.
Figure~\ref{fig5} reveals that the kinetic energy of the star, $T$,
continues to decrease for about two rotation periods after the rate
of emission of $J$ into GR falls to zero.

\bfig[htb]
\centerline{\psfig{file=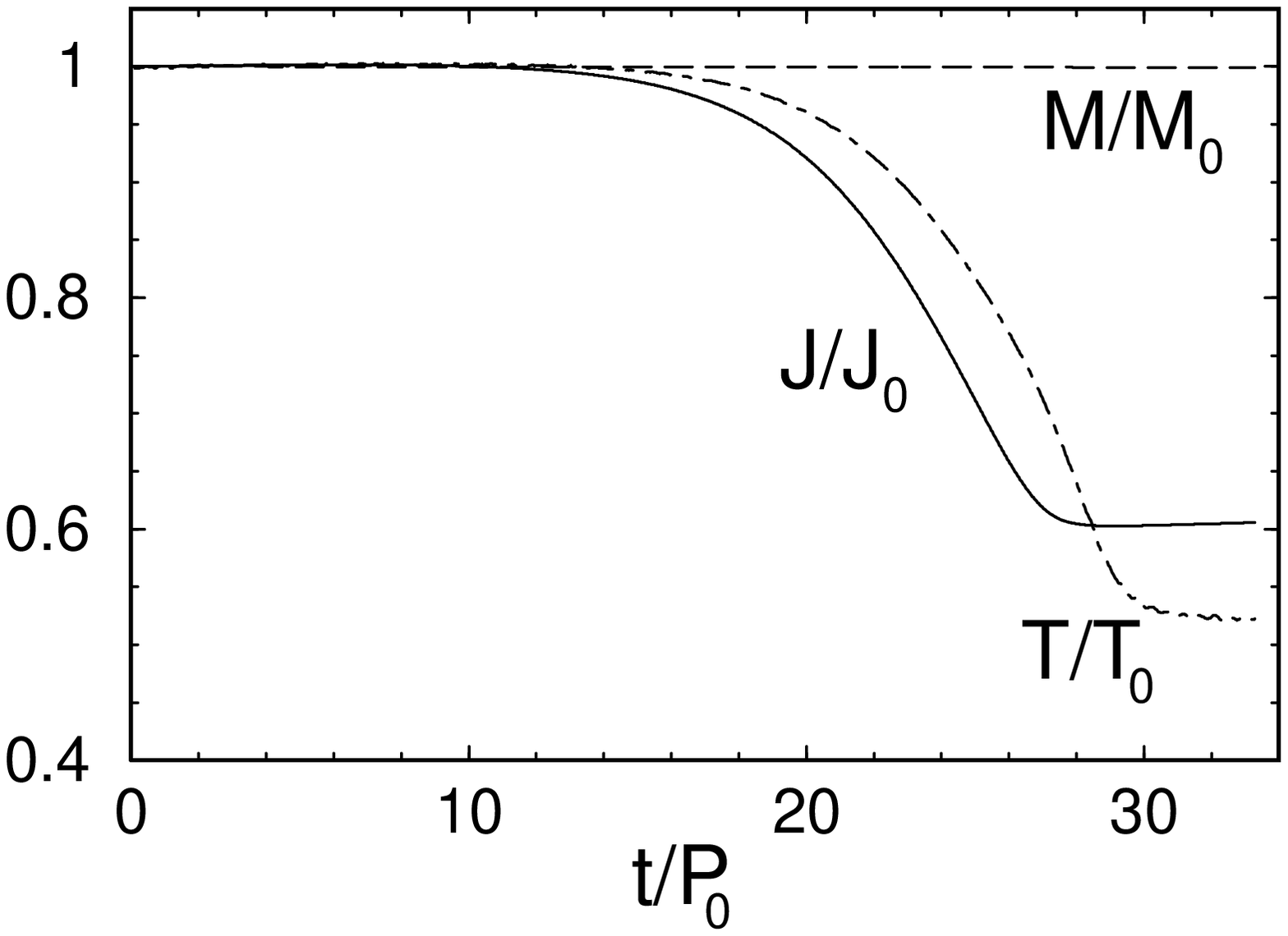,height=1.6in}}
\vskip 0.1cm
\caption{Evolution of the total angular momentum $J/J_0$ (solid curve),
mass $M/M_0$ (dashed), and kinetic energy $T/T_0$ (dot-dashed) of the star.
\label{fig5}}
\efig

What nonlinear process is responsible for limiting the growth of the
$r$-mode?  Figure~\ref{fig5} reveals that the kinetic energy of the
star, $T$, continues to decrease even after the rate of emission
of $J$ into GR falls to zero.  This implies that the energy stored in
the $r$-mode is {\it not} being radiated away as GR or being
transferred to other macroscopic modes.  For if it were, $T$ would be
more or less conserved once the GR losses become small.  Rather we see
that at $t\approx 26P_0$, as the amplitude of the $r$-mode peaks, the
predicted evolution of the total energy $E$ due to GR loss (the dashed
curve in Fig.~\ref{fig6}) significantly diverges from the numerical
evolution of $E$.  This nonconservation of $E$ is due to the
formation of shocks associated with the breaking of surface waves as
illustrated in Fig.~\ref{fig7}~\cite{note4}.

\bfig[htb]
\centerline{\psfig{file=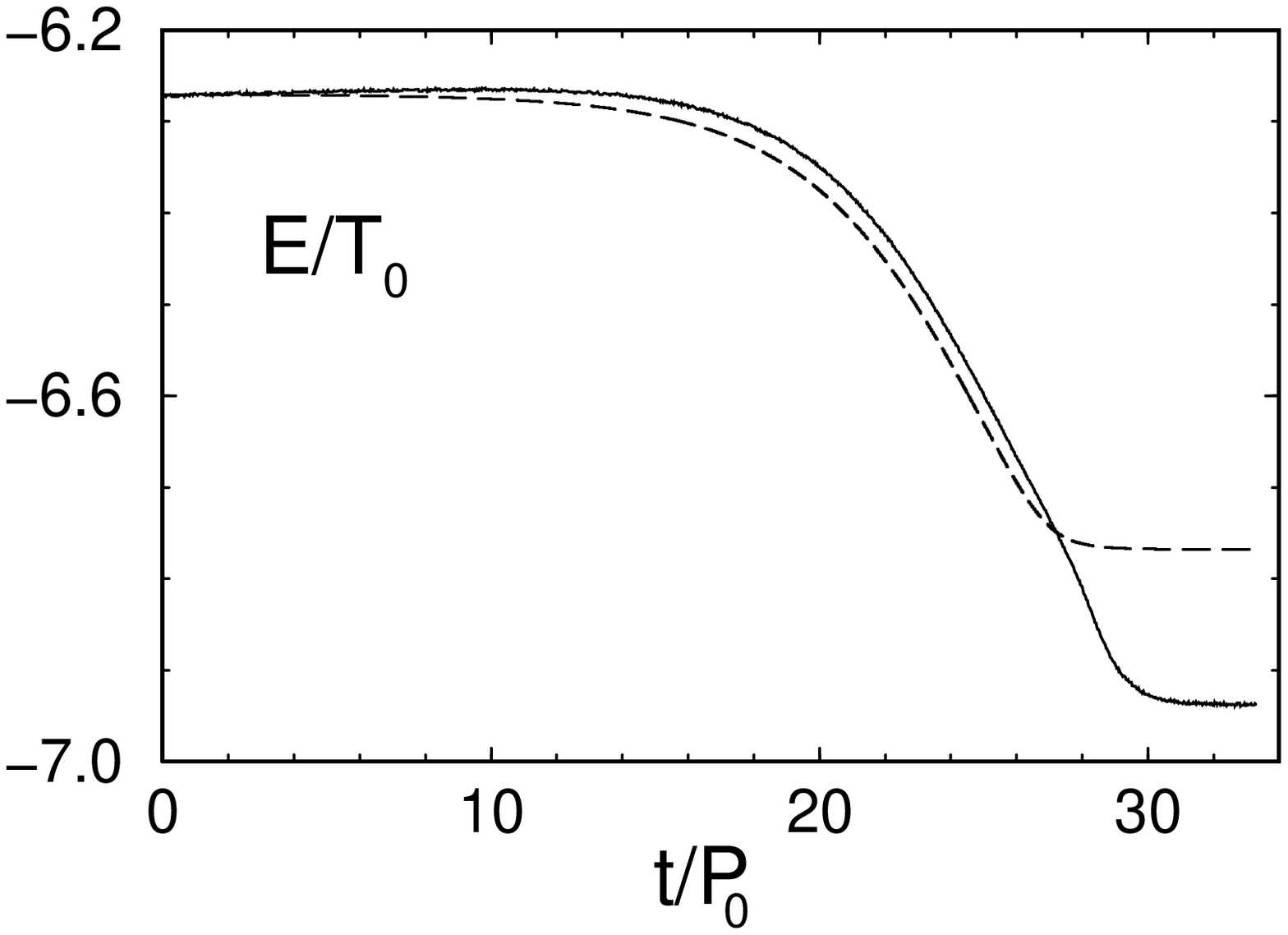,height=1.6in}}
\vskip 0.2cm
\caption{Evolution of the total energy $E$ divided by the initial
rotational kinetic energy $T_0$.  Solid curve is the actual numerical
evolution; dashed curve is the predicted evolution due to GR losses
alone using Eq.~(\ref{eq11}).
\label{fig6}}
\efig

In summary then, we find that an $r$-mode can grow to relatively large
amplitude under the influence of the GR reaction force.  The
hydrodynamic mechanism that acts to suppress the $r$-mode in our
simulation is the formation of shocks near the surface of the star.
Since the GR reaction force in our simulation is too strong by a
factor of 4500, it is still possible that slower hydrodynamic
processes (like the transfer of energy to other modes) could limit the
$r$-mode at smaller values of $\alpha$.  It is also possible that the
coupling of the $r$-mode to $g$-modes in real neutron-star matter (but
absent from our barotropic simulation) could also limit the growth at
smaller $\alpha$.  However if shock formation turns out to be the
dominant suppression mechanism, then we expect the peak amplitude to
be relatively insensitive to the strength of the GR
coupling.~\cite{note5} In this case the dimensionless amplitude of the
mode will peak at about $\alpha=3.4$.  This implies that the 
spindown the star by GR will occur in about one tenth
($\propto1/\alpha^2_{\rm max}$) the time previously
estimated~\cite{lom}.

\bfig[htb]
\centerline{\psfig{file=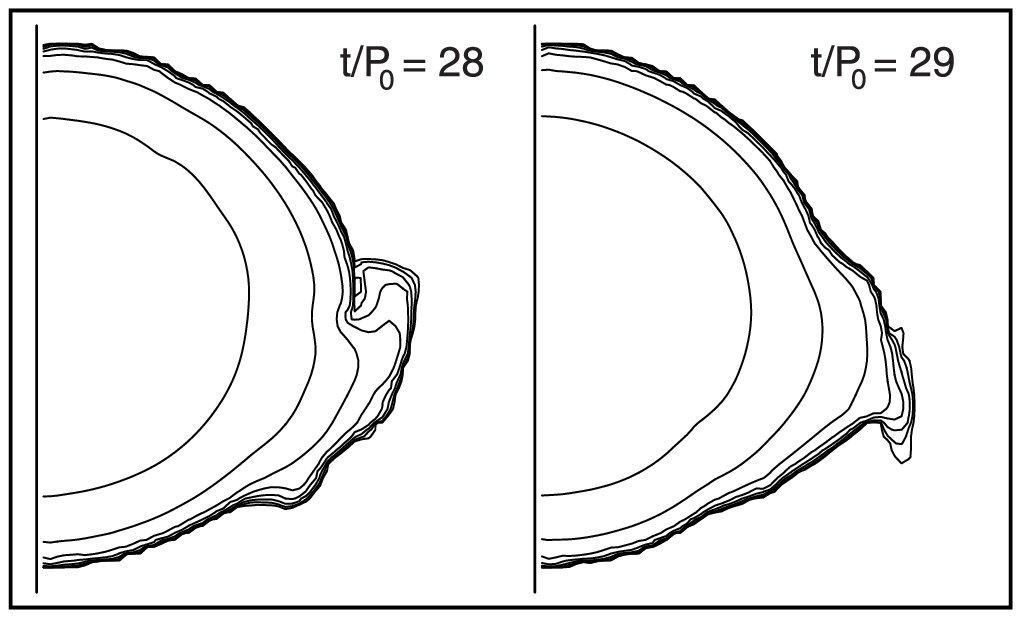,height=1.6in}}
\vskip 0.5cm
\caption{Density contours (at $10^{-n/2}\rho_{\rm max}$ with
$n=1,2,...$) in selected 
meridional planes at times $t=28P_0$ and $29P_0$ illustrate the
breaking of surface waves. Shocks at the leading edges of these waves
appear to be the primary mechanism that suppresses the $r$-mode.
\label{fig7}}
\efig

During our simulation about 40\% of the initial angular momentum and
50\% of the initial rotational kinetic energy is radiated away as GR.
Most of this energy is radiated in a much narrower frequency band
$\Delta f \approx 0.05 f$ than had been expected~\cite{owen}.
Further, the frequency of the $r$-mode is about 20\% smaller than that
predicted by simple perturbation theory.  Both of these effects tend
to make the GR emitted by this process more easily detectable by LIGO.
Even at the end of our simulation the star is still rather rapidly
rotating, and we presume that GR will again drive the $r$-mode
unstable and a second episode of spindown will occur.  During our
simulation about 16\% of the initial rotational kinetic energy of the
star is dissipated by the shocks.  This energy will be converted to
heat in a real neutron star and this would delay cooling and the
formation of a crust.  A detailed thermal analysis will have to be
carried out to determine exactly what other effects this thermal
energy may have.

\acknowledgments

We thank J. Friedman, B. Owen, K. Thorne, and R. Wagoner for helpful
discussions. We also thank H. Cohl, J. Cazes, and especially P. Motl
for contributions to the LSU hydrodynamic code.  This research was
supported by NSF grants PHY-9796079, AST-9987344, AST-9731698,
PHY-9900776, and PHY-9907949, and NASA grants NAG5-4093 and NAG5-8497.
We thank NRAC for computing time on NPACI facilities at SDSC where
tests were conducted; and we thank CACR for access to the HP V2500
computers at Caltech, where the primary simulations were performed.

%%%%%%%%%%%%%%%%%%%%%%%%%%%%%%%%%%%%%%%%%%%%%%%%%%%%%%%%%%%%%%%%%%%%%%%%%%%%%%


\begin{references}
%%%%%%%%%%%%%%%%%%%%%%%%%%%%%%%%%%%%%%%%%%%%%%%%%%%%%%%%%%%%%%%%%%%%%%%%%%%%%%

\bibitem{andersson} N.~Andersson, \apj {\bf 502}, 708 (1998).

\bibitem{fm} J.~L.~Friedman, and S.~M.~Morsink, \apj {\bf 502}, 714 (1998).

\bibitem{lom} L.~Lindblom, B.~J.~Owen, S.~M.~Morsink, \prl {\bf 80},
4843 (1998).

\bibitem{owen} B.~J.~Owen, L.~Lindblom, C.~Cutler, B.~F.~Schutz, A.~Vecchio,
and N.~Andersson, \prd {\bf 58}, 084020 (1998).

\bibitem{stergioulas} N.~Stergioulas, and J.~A.~Font, gr-qc/0007086,
evolve relativistic fluids in the Cowling approximation without
GR reaction and find no $r$-mode saturation at unit amplitude.

\bibitem{blanchet} L.~Blanchet, \prd {\bf 55}, 714 (1997).

\bibitem{rezzolla} L.~Rezzolla, M.~Shibata, H.~Asada, T.~W.~Baumgarte, 
and S.~L.~Shapiro, \apj {\bf 525}, 935 (1999).

\bibitem{tohline} K.~C.~B.~New, and J.~E.~Tohline, \apj {\bf 490}, 311 (1997);
                  H.~S.~Cohl, and J.~E.~Tohline, \apj {\bf 527}, 527 (1999);
                  J.~E.~Cazes, and J.~E.~Tohline, \apj {\bf 532}, 1051 (2000).
 
\bibitem{stone_norman} J.~M.~Stone, and M.~L.~Norman, \apj Supp. {\bf 80},
                       753 (1992).      

\bibitem{note1}   We have verified that this definition
of $\omega$, Eq.~(\ref{eq9}), agrees well with other (numerically
less stable) definitions such as the time derivative of the
phase of $J_{22}$.

\bibitem{diffrot} H.~Spruit, Astron. Astrophys. {\bf 341}, L1 (1999); 
L.~Rezzolla, F.~K.~Lamb, and S.~L.~Shapiro, \apj {\bf 531}, L139
(2000); Y.~Levin, and G.~Ushomirsky, Mon. Not. Roy. Astr. Soc. in
press (2000); W.~C.~G.~Ho, and D.~Lai, \apj {\bf 544}, in press
(2000).

\bibitem{thorne} K.~S.~Thorne, Rev. Mod. Phys. {\bf 52}, 299 (1980).

\bibitem{note3} The difference between the actual and the predicted 
evolution of $E$ (at early times) and $J$ was due primarily to
numerical error introduced by the Coriolis term in the rotating-frame
evolution equations.  A higher-order expression for this term has now
been implemented, and even better agreement is expected in future
simulations.

\bibitem{note4}  In the barotropic hydrodynamic code
used in this simulation, the structure of a shock is correctly
evolved, but the energy dissipated in the shock is simply ignored.

\bibitem{note5} We have verified that $\alpha_{\rm max}$ is
insensitive to the value of $\kappa$ by running an additional
simulation. We find that $\alpha_{\rm max}$ decreases by 0.36\%
when $\kappa$ is increased by the factor 4/3.

\end{references}
\end{document}